\documentclass[aps,prd,reprint,showpacs,floatfix,nofootinbib,superscriptaddress]{revtex4-1}
\usepackage{color,amsmath,amssymb,amsthm,graphicx,latexsym}
\usepackage{graphics,bm}
\usepackage{amssymb}
\usepackage{array}
\usepackage{mathtools}
\usepackage{psfrag}

\usepackage{amsthm}
\usepackage{newlfont}
\usepackage{esdiff}
\usepackage{multirow}
\usepackage{graphics}
\usepackage[english]{babel}
\usepackage{graphicx}
\usepackage{ae,aecompl,color}
\usepackage{bm}
\usepackage{booktabs}
\usepackage[bookmarks=true,bookmarksopen=true,bookmarksnumbered=true,bookmarksopenlevel=3]{hyperref}

\makeatletter
\DeclareRobustCommand*{\bfseries}{%
  \not@math@alphabet\bfseries\mathbf
  \fontseries\bfdefault\selectfont
  \boldmath
}
\makeatother

\definecolor{airforceblue}{rgb}{0.36, 0.54, 0.66}
\definecolor{steelblue}{rgb}{0.27, 0.51, 0.71}
\definecolor{amber}{rgb}{1.0, 0.49, 0.0}

\begin{document}
\newcommand{\bd}{\begin{document}}
\newcommand{\ed}{\end{document}}
\newcommand{\bc}{\begin{center}}
\newcommand{\ec}{\end{center}}
\newcommand{\bfr}{\begin{flushright}}
\newcommand{\efr}{\end{flushright}}
\newcommand{\lt}{\left}
\newcommand{\rt}{\right}
\newcommand{\vs}{\vspace}
\newcommand{\hs}{\hspace}
\newcommand{\beq}{\begin{equation}}
\newcommand{\eeq}{\end{equation}}
\newcommand{\lb}{\linebreak}
\newcommand{\pb}{\pagebreak}
\newcommand{\mb}{\makebox}
\newcommand{\fb}{\framebox}
\newcommand{\mc}{\multicolumn}
\newcommand{\ben}{\begin{enumerate}}
\newcommand{\een}{\end{enumerate}}
\newcommand{\bit}{\begin{itemize}}
\newcommand{\eit}{\end{itemize}}
\newcommand{\ovl}{\overline}
\newcommand{\un}{\underline}
\newcommand{\lefq}{\lefteqn}
\newcommand{\ba}{\begin{array}}
\newcommand{\ea}{\end{array}}
\newcommand{\beqa}{\begin{eqnarray}}
\newcommand{\eeqa}{\end{eqnarray}}
\newcommand{\beqas}{\begin{eqnarray*}}
\newcommand{\eeqas}{\end{eqnarray*}}
\newcommand{\bfg}{\begin{figure}}
\newcommand{\efg}{\end{figure}}
\newcommand{\bds}{\begin{displaymath}}
\newcommand{\eds}{\end{displaymath}}
\newcommand{\btb}{\begin{tabbing}}
\newcommand{\etb}{\end{tabbing}}
\newcommand{\para}{\parallel}
\newcommand{\pad}{\partial}
\newcommand{\nn}{\nonumber}
\newcommand{\la}{\leftarrow}
\newcommand{\ra}{\rightarrow}
\newcommand{\lgla}{\longleftarrow}
\newcommand{\lgra}{\longrightarrow}
\newcommand{\La}{\Leftarrow}\newcommand{\Ra}{\Rightarrow}
\newcommand{\Lra}{\Leftrightarrow}
\newcommand{\Lgla}{\Longleftarrow}
\newcommand{\Lgra}{\Longrightarrow}
\newcommand{\lan}{\langle}
\newcommand{\ran}{\rangle}
\renewcommand{\a}{\alpha}
\renewcommand{\b}{\beta}
\newcommand{\g}{\gamma}
\newcommand{\G}{\Gamma}
\renewcommand{\d}{\delta}
\newcommand{\eps}{\epsilon}
\newcommand{\s}{\sigma}
\newcommand{\D}{\Delta}
\newcommand{\vare}{\varepsilon}
\newcommand{\pr}{\prime}
\newcommand{\ro}{\rho}
\newcommand{\nab}{\nabla}
\newcommand{\m}{\mu}
\newcommand{\n}{\nu}
\newcommand{\Sg}{\Sigma}
\newcommand{\p}{\pi}
\newcommand{\R}{I\!\!R}
\newcommand{\om}{\omega}
\newcommand{\Om}{\Omega}
\newcommand{\ze}{\zeta}
\newcommand{\vart}{\vartheta}
\newcommand{\lam}{\lambda}
\newcommand{\tri}{\triangle}
\newcommand{\f}{\frac}
\newcommand{\iny}{\infty}
\newcommand{\pro}{\propto}
\newcommand{\np}{\newpage}

\title{Quantum phase transitions  of the Dirac oscillator in the Anti-Snyder model} 
%\vs{1cm}
%
%
\author{\textsc{M.~Presilla}}
\affiliation{Dipartimento di Fisica e Geologia, Universit\`a degli Studi di Perugia, Via A.~Pascoli, I-06123 Perugia, Italy}
%\altaffiliation[Current Address: ]{Dipartimento di Fisica, Universit\`a di Bologna}
\author{\textsc{O.~Panella}}
\affiliation{Istituto Nazionale di Fisica Nucleare, Sezione di Perugia, Via A.~Pascoli, I-06123 Perugia, Italy}
\email[({\bf Corresponding Author})\\ Email: ]{orlando.panella@pg.infn.it }

\author{\textsc{P.~Roy}}
\affiliation{Physics and Applied Mathematics Unit, Indian Statistical Institute, Kolkata-700108, India}

\date{\today}

\begin{abstract}
We obtain exact solutions of the (2+1) dimensional Dirac oscillator in a homogeneous  magnetic field within the Anti-Snyder modified uncertainty relation characterized by a momentum cut-off ($p\leq p_{\text{max}}=1/ \sqrt{\beta}$). In ordinary quantum mechanics ($\beta\to 0$) this system is known to have a single left-right chiral  quantum phase transition (QPT). 
We show that a finite momentum cut-off modifies the spectrum introducing additional quantum phase transitions. It is also shown that the presence of momentum cut-off modifies the degeneracy of the states.
\end{abstract}

\maketitle
\section{Introduction}
In recent years quantum mechanical models based on modified commutation relations (MCR) have been studied by many authors. Usually MCR's involve a minimum position uncertainty leading to a minimal length~\cite{Kempf:1994su,Kempf:1996fz} ($\hbar\sqrt{\beta}$). But there are also MCR's which lead to a maximum 
momentum~\cite{Magueijo:2001cr,Magueijo:2002am,Cortes:2004qn,AmelinoCamelia:2000ge,Das:2010zf,Pedram:2011zz,Maggiore:1993kv,Battisti:2009zzb} (${1}/{\sqrt{\beta}}$). A number of quantum mechanical models in the presence of a maximum momentum parameter have also been investigated~\cite{Mignemi:2011gr,Ching:2012vq,Ching:2013jua,Ching:2013lma,Ching:2013ala}. The main reason for studying quantum mechanical models with MCR is that the spectrum gets modified because of the presence of a minimum length/maximum momentum parameter which may eventually be detected in low energy experiments. 

The Dirac oscillator \cite{Ito:1967aa,Cook:1971hp,Moshinsky:1989aa,Benitez:1990te,Rozmej:1999jv,Sadurni:2009wk} in the presence of a homogeneous magnetic field is one of the few exactly solvable problems in relativistic quantum mechanics \cite{Mandal:2009we}. This combined system is known to exhibit a quantum chirality phase transition whenever the magnetic field strength exceeds a critical 
value~\cite{Quimbay:2013aa,Hou:2014aa,Bermudez:2007ab}. 
Such system has received quite a lot of attention also in view of the possible applications to the physics of recently discovered materials like graphene~\cite{Geim:2007aa,Castro-Neto:2009aa,Semenoff:1984aa}, germanene~\cite{Davila:2014aa,Cahangirov:2009aa} and silicene~\cite{Lalmi:2010kx,Vogt:2012vn,Fleurence:2012ys,Lin:2012zr}. Indeed it has been known that, in these materials, at the $K$, $K'$ points of the Brillouin zone the charge carriers are described by an effective 2 dimensional Dirac equation. The Dirac oscillator coupling  has been proposed~\cite{Quimbay:2013ab} to arise from the interaction of the charge carriers with an effective \emph{internal} magnetic field due to collective motion of the electrons in the planar hexagonal lattice of the carbon atoms.

This system has also been studied in the presence of a minimal length \cite{Menculini:2014isa} as well as in non-commutative space \cite{Panella:2014hga,Hou:2014oqa}. In all these cases it was shown that there are a multitude of quantum phase transitions depending on the minimal length/non-commutative parameters. In this work our objective is to study again the same system but in the Anti-Snyder space i.e, under a MCR which incorporates a momentum cut-off. One reason why we have taken up this model is that in contrast to the minimum length scenario in the present case the presence of a maximum momentum parameter changes the nature of the spectrum from an infinite one to a finite one. Consequently it is of interest to examine how the quantum phase transitions are influenced by the maximum momentum parameter. 

We will show that in the Anti-Snyder model the (2+1)-dimensional Dirac oscillator of frequency $\omega$ in an external magnetic field ($B_0$) admits a finite and bounded spectrum (of bound states). 
Similarly to what has been found in ref.~\cite{Menculini:2014isa} we find that the excited energy levels can be classified in two classes one of which disappears in the ordinary quantum mechanical limit, $\beta \to 0$, and whose corresponding  states are characterised by the presence of quantum phase transitions which accumulate towards the  critical field $B_0=\frac{2\omega Mc}{e}$.  The other class of levels does not disappear in the ordinary quantum mechanical limit, $\beta \to 0$, and indeed reduces to the known spectrum of the system in this limit.
We postpone the discussion of the continuum states to a future work.   

We would also like to recall that recently the one dimensional version of the Dirac oscillator has been realised in the laboratory~\cite{Franco-Villafane:2013aa}. In addition practicable prospects of realising soon the two dimensional version of the Dirac oscillator have  also been reported in the literature~\cite{Franco-Villafane:2013aa,Sadurni:2010fk,Hul:2005uq}.
The exact solutions presented here may have then a direct relevance to the physics of experimentally accessible 2-dimensional systems which have received a lot of attention recently in the literature like~\cite{Geim:2007aa,Castro-Neto:2009aa,Semenoff:1984aa,Lalmi:2010kx,Vogt:2012vn,Fleurence:2012ys,Lin:2012zr,Davila:2014aa,Cahangirov:2009aa}.  

The rest of the paper is organised as follows:
Sec.~\ref{section:formulation} introduces the Anti-Snyder model and  defines the problem; Sec.~\ref{section:transformation} describes the transformation of the problem to a P\"{o}schl-Teller potential;
Sec.~\ref{section:weak} and  Sec.~\ref{section:strong} present the explicit solutions respectively in the case of weak and strong magnetic field.
Sec.~\ref{section:discussion} gives a discussion of the main properties of the spectrum and the quantum phase transitions. Finally Sec.~\ref{section:conclusion} presents the conclusions.

\section{Formulation of the problem}
\label{section:formulation}
Let us first present the basic facts about the Anti-Snyder space. The commutation relations between the coordinates $({\hat x}_i)$ and momenta $({\hat p}_j)$ are given by~\cite{Mignemi:2011gr}:
\begin{equation}
\begin{array}{ll}
[{\hat x}_i,{\hat x}_j]=2i\hbar\b(1-\b{\hat p}^2)\epsilon_{ijk}L_k, & [{\hat p}_i,{\hat p}_j]=0,\cr
[{\hat x}_i,{\hat p}_j]=i\hbar \delta_{ij}(1-\b {\hat p}^2), &[{\hat p}_i,L_j]=i\hbar\epsilon_{ijk}{\hat p}_k, \cr
[{\hat x}_i,L_j]=i\hbar\epsilon_{ijk}{\hat x}_k, &[L_i,L_j]=i\hbar\epsilon_{ijk}L_k,\cr
\end{array}
\end{equation}
where $L_i=\f{1}{1-\b {\hat p}^2}\epsilon_{ijk}{\hat x}_j{\hat p}_k$. A representation of the operators ${\hat x}_i$ and ${\hat p}_i$ can be taken as
\beq
{\hat x}_i=i\hbar(1-\b p^2) \f{\partial}{{\partial p}_i},~~~~{\hat p}_i=p_i
\eeq
It may be pointed out that in the present case the scalar product is defined as
\beq
\langle f|g \rangle =\int_{0}^{\f{1}{\sqrt{\b}}} \f{d^2\bm p}{(1-\b {\bm p}^2)}f^*({\bm p})g({\bm p})\label{scalar}
\eeq
Let us now consider the Hamiltonian corresponding to $(2+1)$ dimensional Dirac oscillator in the presence of a homogeneous magnetic field $\bm{B}= (0,0,B_0)$. In the Anti-Snyder space the corresponding Hamiltonian is given by 
\beq\label{h1}
H=c{\bm\s}\cdot({\hat{{\bm p}}-iM\omega\s_z\hat{\bm x}+\frac{e}{c}\hat{\bm A}})+\s_z Mc^2\, ,
\eeq
where $\bm{\s}=(\s_x,\s_y)$ and $\hat{\bm A}= (-B_0{\hat y}/2,B_0{\hat x}/2,0)$ denote the Pauli matrices and the vector potential respectively. We shall now solve the eigenvalue equation corresponding to the Hamiltonian (\ref{h1}). The eigenvalue problem reads
\beq
 H \psi =
\left(\ba{cc}Mc^2 & cP_-\\cP_+ & -Mc^2\ea\right)\left(\ba{c}\psi^{(1)} \\ \psi^{(2)}\ea\right)=E\left(\ba{c}\psi^{(1)} \\ \psi^{(2)}\ea\right) 
\label{H}
\eeq
or
\begin{subequations}
\label{intertwining}
\begin{align}
\label{intertwining1}cP_-\psi^{(2)} &= \epsilon_{-}\, \psi^{(1)}\\
\label{intertwining2}cP_+\psi^{(1)} &= \epsilon_{+}\, \psi^{(2)}
\end{align}
\end{subequations}
where $\epsilon_{\pm} = E\pm Mc^2$ and  the operators $P_{\pm}$ are given by
\beq
\ba{lcl}
P_+&=& e^{+i\vart}\left[p-\lam\left(1-\b p^2\right)\left(\partial_p+\frac{i}{p}\partial_\vart\right) \right]~, \\
P_-&=& e^{-i\vart}\left[p+\lam\left(1-\b p^2\right)\left(\partial_p-\frac{i}{p}\partial_\vart\right) \right]~, \label{polPmETA}
\ea
\eeq
with 
\beq\label{om}
\lam=\f{\hbar eB_0}{2c}-M\hbar\omega=M\hbar \left(\tilde{\omega}_c-\omega\right), \qquad \tilde{\omega}_{c}=\f{eB_0}{2Mc}
\eeq
\beq
\label{momentumspace}
p_x=p~\cos\vart,~~~~p_y=p~\sin\vart,~~~~p_x^2+p_y^2=p^2
\eeq
From Eq.~\eqref{om} it is seen that depending on the strength of the magnetic field
(in comparison to the oscillator strength), $\lam$ can either
be positive or negative. 
\section{Transformation to a P\"{o}schl-Teller II potential}
\label{section:transformation}

%Let us first consider the case $\lam>0$.  
 In order to solve the eigenvalue problem in Eq.~(\ref{H}) we proceed as follows:
\beq
\psi^{(1)}=e^{im\vart}\varphi^{(1)}_m(p)\,,~~~\psi^{(2)}=e^{i(m+1)\vartheta}\varphi^{(2)}_m(p)
\eeq
where $m=0,\pm 1,\pm 2,...$ denotes the angular momentum quantum number in the $p$ space.

Now applying the operator $H$ from the left on Eq.~\eqref{H}, the eigenvalue equations for the 
 components $\psi^{(1,2)}$ can be found as:
 \begin{subequations}
 \label{secondorder}
 \begin{align}
 \label{secondorder1} P_-P_+\psi^{(1)} &= \epsilon^2 \psi^{(1)}\,,\\
 \label{secondorder2}P_+P_-\psi^{(2)} &= \epsilon^2 \psi^{(2)}\,,\\
 \label{secondorder3}\epsilon^2 &= \frac{E^2-M^2c^4}{c^2}= \frac{\epsilon_{+}\epsilon_{-}}{c^2}\,.
 \end{align}
 \end{subequations}
 With a little algebra then the above Eqs.~(\ref{secondorder1},\ref{secondorder2}) can be put in the following form:
\beq\label{sch1}
- f(p)\f{d^2\varphi^{(1)}}{dp^2}+g(p)\f{d\varphi^{(1)}}{dp}+h_1(p)\varphi^{(1)}=(\eps^2/\b\lam^2)\varphi^{(1)}
\eeq
\beq\label{sch2}
- f(p)\f{d^2\varphi^{(2)}}{dp^2}+g(p)\f{d\varphi^{(2)}}{dp}+h_2(p)\varphi^{(2)}=(\eps^2/\b\lam^2)\varphi^{(2)}
\eeq
where 
\begin{eqnarray}
f(p)&=&\f{(1-\b p^2)^2}{\b}\nonumber\\
g(p)&=&\left(1-\beta p^2\right)\left[2p-\dfrac{\left(1-\beta p^2\right)}{\beta p}\right]\nonumber\\
h_1(p)&=&\dfrac{p^2}{\beta \lambda^2} +\dfrac{(1-\beta p^2)(2m+2-2m\beta \lambda)}{\beta \lambda} \nonumber \\ &&\phantom{xxxxxxxxxxxx}+\dfrac{(1-\beta p^2)^2 m^2}{\beta p^2}\\
h_2(p)&=&\f{p^2}{\b\lam^2}+(2m+2m\b\lam +2 \b\lam)\f{(1-\b p^2)}{\b\lam}\nonumber \\
&&\phantom{xxx}+\f{(m+1)^2(1-\b p^2)^2}{\b p^2}
\end{eqnarray}

In order to solve the equations (\ref{sch1}) and (\ref{sch2}) we perform the transformations
\beq
\varphi^{(1,2)}(p) = \rho(p)\phi_{1,2}(p),~~~~q = \int \f{1}{\sqrt{f(p)}}~dp\label{t}
\eeq
\beq
\rho(p) = e^{\int \chi(p)~dp},~~~~\chi(p) =  \f{f^\prime+2g}{4f}\label{rho}
\eeq
namely
\beq
\varphi^{(1,2)}(p) = p^{-1/2} \phi_{1,2}(p),~~~~q = \tanh^{-1}(\sqrt{\b} p)
\eeq

%we obtain from (\ref{sch2})
%\beq
%\left[-\f{d^2}{dq^2} + V_2(q)\right]\phi_2(q) = (\eps^2/\b\lam^2)\phi_2(q)\label{sch3}
%\eeq
%The potential $V_2(q)$ is given by
%\beq
%V_2(q) = \left[\f{4g_2^2+3{f_2^\prime}^2+8g_2f_2^\prime}{16f_2}-\f{f_2^{\prime\prime}}{4}-\f{g_2^\prime}{2} + h_2(p)\right]_q
%\eeq
%and in terms of the new variable $q$ it reads
%\beq
%V_2(q)=-\f{\b\lam(4m^2\b\lam-8m-\b\lam)-4}{4\b^2\lam^2}~sech^2q+\f{4m^2+8m+3}{4}~cosech^2q+\f{1}{\b^2\lam^2}
%\eeq
and obtain from Eqs.(\ref{sch1}) and (\ref{sch2})
\begin{subequations}
\label{effSch}
\begin{align}
\label{effSch1}  
\left[-\f{d^2}{dq^2} + V_i(q)\right]\phi_i(q) \,=\,\,& \,k^2\phi_i(q),~~~~i=1,2\\&k^2=\f{\eps^2-1/\b}{\b\lam^2}\,.\label{reducedeigenvalues}
\end{align}
\end{subequations}
Eq.~(\ref{effSch1}) is effectively an one dimensional Schr\"odinger equation with a potential given by
\begin{eqnarray}
\label{pot1}
V_1(q)&=&-\f{(-2+\b\lam+2m\b\lam)(-2+3\b\lam+2m\b\lam)}{4\b^2\lam^2}~\text{sech}^2q\nonumber \\&&+(m-\f{1}{2})(m+\f{1}{2})~\text{cosech}^2q,~~0<q<\infty\\
\label{pot2}
V_2(q)&=&-\f{(-2-\b\lam+2m\b\lam)(-2+\b\lam+2m\b\lam)}{4\b^2\lam^2}~\text{sech}^2q\nonumber\\ &&+(m+\f{1}{2})(m+\f{3}{2})~\text{cosech}^2q,~~0<q<\infty
\end{eqnarray}
It may be noted that the potentials (\ref{pot1}) and (\ref{pot2}) can be identified with the P\"{o}schl-Teller potential II of the form \cite{Cooper:525919,Dabrowska:1987fd}
\begin{equation}
\begin{array}{l}
\label{pt2}
U(x)=-A(A+1)~\text{sech}^2x + B(B-1)~\text{cosech}^2x, \\
0<x<\infty, \qquad A>B>0
\end{array}
\end{equation}
The above potential is exactly solvable. The eigenvalues and the corresponding (unnormalised) eigenfunctions of the Schr\"odinger eigenvalue problem as in Eq.~(\eqref{effSch1}) are given by \cite{Cooper:525919,Dabrowska:1987fd}:
\begin{subequations}
\label{sol}
\begin{align}
\label{sol1}k^2_n&=-(A-B-2n)^2, ~~~ n=0,1,2,...<\f{A-B}{2}\,,\\ 
\begin{split}
%\phi_{n} &= N (\cosh 2x - 1)^{B/2} (\cosh 2x +1)^{-A/2} P_{n}^{(B-1/2,-A-1/2)}(\cosh 2x) 
\phi_{n}&=N (\cosh 2x - 1)^{B/2} (\cosh 2x +1)^{-A/2}\times \\ &\phantom{xxx}{\phantom{F}}_{2}F_{1}(-n,B-A+n;B+1/2; \tfrac{1- \cosh 2x}{2})
\end{split}
%\phi_n&=(y-1)^{B/2}(y+1)^{-A/2}~P_n^{(B-1/2,-A-1/2)}(y)\,,
%\\y&=2\,\text{sinh}^2x+1\,,
\end{align}
\end{subequations}
%where $P_n^{\alpha,\beta}(y)$ are Jacobi polynomials. 
where $\!\!\!\!{\phantom{F}}_{2}F_{1}(-n,B-A+n;B+1/2; \tfrac{1- \cosh 2x}{2})$ is the hyper-geometric function, and $N$ a normalisation constant.
Note that the conditions $A>B>0$ are required for the wave functions to be normalizable. Now identifying the potential (\ref{pot2}) with the one in (\ref{pt2}) we have to solve the following two systems of quadratic equations in the unknown quantities ($A_1,B_1$) and ($A_2,B_2$):
\begin{eqnarray}
A_1(A_1+1)&=&\f{(-2+\b\lam+2m\b\lam)(-2+3\b\lam+2m\b\lam)}{4\b^2\lam^2}\,,\nonumber\\
B_1(B_1-1)&=&(m-\f{1}{2})(m+\f{1}{2})\,;\nonumber\\
A_2(A_2+1)&=&\f{(-2-\b\lam+2m\b\lam)(-2+\b\lam+2m\b\lam)}{4\b^2\lam^2}\,,\nonumber\\
B_2(B_2-1)&=&(m+\f{1}{2})(m+\f{3}{2}).\nonumber
\end{eqnarray}
Each system admits four solutions ($a,b,c,d$) which are shown in Table~\ref{combination1} and Table~\ref{combination2} (first and second columns). In order to ensure that the resulting wave functions are normalizable it is necessary to impose the conditions $A_j>B_j>0$ for $j=1,2$ whose resulting constraints are shown in the third and fourth columns of Table~\ref{combination1} and Table~\ref{combination2}. 
%\beq\label{combination}
%\ba{l}
%\displaystyle A=\f{1}{\b\lam}-\f{1}{2}-m,~~~~B=-m-\f{1}{2}\\
%\displaystyle A=\f{1}{\b\lam}-\f{1}{2}-m,~~~~B=m+\f{3}{2}\\
%\displaystyle A=-\f{1}{\b\lam}-\f{1}{2}+m,~~~~B=-m-\f{1}{2}\\
%\displaystyle A=-\f{1}{\b\lam}-\f{1}{2}+m,~~~~~B=m+\f{3}{2}
%\ea
%\eeq
%%%%%%%%%%%%%%%%%%%%%%%%%%%%%%%%%%%%%%%%%%%%%%%%%%%%%%%%%%%%%%%%%%%%%%%
%V2%%%%%%%%%
\begin{table*}
\begin{ruledtabular} 
\begin{tabular}{lccccccc}
& $A_1$
& $B_1$ 
& $A_1>B_1$
& $B_1>0$&$\lambda <0$&$ 0<\lambda\,\,  < \frac{1}{2\beta}$&$k_{n,m}^2$\\
\hline
$(a)$
& $  -\f{1}{\b\lam}+\f{1}{2}+m$
&$-m+\f{1}{2} $
& $m>\f{1}{2\b\lam} $&$m\leq 0$&$\frac{1}{2\beta\lambda} <m\leq0$&--&$-4 \left(-\frac{1}{2\beta\lambda}+m -n\right)^2$\\
\hline
$(b)$
& $-\f{1}{\b\lam}+\f{1}{2}+m  $
& $ m+\f{1}{2} $
& $ -\f{1}{\b\lam}>0$&$m \geq 0$&$m\geq 0$&--& $ -4\left(-\frac{1}{2\beta\lambda}-n\right)^2$\\
\hline
$(c)$ 
& $\f{1}{\b\lam}-\f{3}{2}-m$
& $-m+\f{1}{2}$ 
& $ \f{1}{\b\lam}>2$&$m\leq 0$&--&$m\leq 0$&$-4\left(\frac{1}{2\beta\lambda}-1-n\right)^2$\\
\hline
$(d)$
& $\f{1}{\b\lam}-\f{3}{2}-m $
& $m+\f{1}{2}$
& $m< \frac{1}{2\beta\lambda}-1$&$m\geq 0$ &  --&$0\leq m<\frac{1}{2\beta\lambda}-1$&$-4\left(\frac{1}{2\beta\lambda}-1-m-n\right)^2$\\
\end{tabular}
\end{ruledtabular}
\caption{\label{combination1} 
We give here the possible solutions $(a,b,c,d)$ of the coefficients $A_1$ and $B_1$ in terms of the parameters  $m$ and $\beta\lambda$ when identifying the potential $V_1(q)$ in Eq.~\ref{pot2} with the P\"{o}schl-Teller potential of type II, $U(x)$ in Eq.~\ref{pt2}. We obtain the $m$ ranges respectively for the cases $\lambda<0$ (fifth column) and $\lambda>0$ (sixth column) combining the results form the conditions $A_1>B_1$ (third column) and $B_1>0$ (fourth column). Cases $(c)$ and $(d)$ when $\lambda >0$ both imply the condition $ \lambda<\frac{1}{2\beta}$.}
\end{table*}
\begin{table*}
\begin{ruledtabular}
\begin{tabular}{lccccccc}
& $A_2$
& $B_2$ 
& $A_2>B_2$&$B_2>0$&$  -\frac{1}{2\beta} <\,  \lambda <0$&$\lambda >0$&$k_{n,m}^2$\\
%& $A>0$
%& $B>0$
\hline
$(a)$
& $  -\f{1}{\b\lam}-\f{1}{2}+m$
&$ -m-\f{1}{2} $
& $m >\f{1}{2\b\lam}$&$m\leq -1$&$\frac{1}{2\beta \lambda} < m \leq -1$ & --&$-4 \left(-\frac{1}{2\beta\lambda}+m -n\right)^2$\\
%& $ m>\f{1}{\b\l}+\f{1}{2}$
%& $ $\\
\hline
$(b)$
& $-\f{1}{\b\lam}-\f{1}{2}+m  $
& $ m+\f{3}{2} $
& $ \frac{1}{\beta \lambda} < -2$&$m\geq -1$&$m\geq -1$ &--&$-4 \left(-\frac{1}{2\beta\lambda}-1 -n\right)^2$\\
\hline
$(c)$ 
& $\f{1}{\b\lam}-\f{1}{2}-m$
& $ -m-\f{1}{2}$ 
& $ \f{1}{\b\lam}>0$ &$m\leq -1$& -- &$m\leq -1$&$-4\left(\frac{1}{2\beta\lambda}-n\right)^2$\\
%& $m<\f{1}{\b\l}-\f{1}{2} $
%& $m<-\f{1}{2} $\\
\hline
$(d)$
& $\f{1}{\b\lam}-\f{1}{2}-m $
& $m+\f{3}{2}$
& $ m < \frac{1}{2\beta \lambda} -1 $&$m\geq -1$& -- & $-1 \leq m < \frac{1}{2\beta \lambda} -1$&$-4\left(\frac{1}{2\beta\lambda}-1-m-n\right)^2$\\
%& $m<\f{1}{\b\l}-\f{1}{2} $
%& $m>-\f{3}{2} $ \\
%& $ $
% $ $\\
\end{tabular}
\end{ruledtabular}
\caption{\label{combination2} The same as in Table~\ref{combination1} but for the potential  $V_2(q)$ in Eq.~\ref{pot1}. Cases $(a)$ and $(b)$ when $\lambda <0$ both imply the condition $ -\frac{1}{2\beta} < \lambda$.  Note that when combining the upper and lower solutions the possible values of  $\lambda$ are  limited to the interval $-\frac{1}{2\beta} <\lambda <\frac{1}{2\beta}$. This can be qualitatively understood since  $\lambda$ defined in Eq.~\ref{om} has the dimension of a squared momentum. Hence, being the Anti-Snyder model characterised by a maximum momentum $p_{max}=1/\sqrt{\beta}$, it follows a bounded allowed range in $\lambda$.   }
\end{table*}
%%%%%%%%%%%%%%%%%%%%%%%%%%%%%%%%%%%%%%%%%%%%%%%%%%%%%%%%%%%%%%%%%%%%%%%
%It is now necessary to examine which of these combinations are compatible with the constraints $A>B>0$. The condition $B<A$ gives
%\beq\label{b<a}
%\ba{l}
%\displaystyle\f{1}{\b\lam}>0\\
%\displaystyle\f{1}{\b\lam}-2-2m>0\\
%\displaystyle-\f{1}{\b\lam}+2m>0\\
%\displaystyle-\f{1}{\b\lam}-2>0
%\ea
%\eeq
%The energy levels of the Dirac equation (corresponding to the different combinations in (\ref{combination})) are then given by
%\beq\label{e}
%\ba{l}
%E_n=\pm c\sqrt{M^2c^2+\f{1}{\b}-4\b\lam^2(\f{1}{2\b\lam}-n)^2},~~~~n=0,1,....<\f{1}{2\b\lam}\\
%E_n=\pm c\sqrt{M^2c^2+\f{1}{\b}-4\b\lam^2(\f{1}{2\b\lam}-1-m-n)^2},~~~~n=0,1,....<\f{1}{2\b\lam}-1-m\\  
%E_n=\pm c\sqrt{M^2c^2+\f{1}{\b}-4\b\lam^2(\f{1}{2\b\lam}-m+n)^2},~~~~n=0,1,....<-\f{1}{2\b\lam}+m\\
%E_n=\pm c\sqrt{M^2c^2+\f{1}{\b}-4\b\lam^2(\f{1}{2\b\lam}+1+n)^2},~~~~n=0,1,....<-\f{1}{2\b\lam}-1\\
%\ea
%\eeq

Before discussing the explicit general solutions for the eigenvalues and wavefunctions based on the above results we conclude this section by a separate discussion of the zero modes, e.g. the states with energy $E=\pm Mc^2$ (or $E=0$ in the massless case). Such states can be identified solving explicitly the first order equations in Eq.~(\ref{H}) which read:
\begin{subequations}
\label{firstorder}
\begin{align}
\label{firstorder1}cP_-\psi^{(2)} &= (E-Mc^2)\, \psi^{(1)}\\
\label{firstorder2}cP_+\psi^{(1)} &= (E+Mc^2)\, \psi^{(2)}
\end{align}
\end{subequations}
From Eqs.~(\ref{firstorder}) it is seen that the eigenfunctions corresponding to the eigenvalues $E=\pm Mc^2$ are correctly indentified by the following set of conditions:
\begin{subequations}
\label{zeromodes}
\begin{align}
\label{zeromodes1}E=+Mc^2, \quad\to\quad  &\psi^{(2)}=0, ~~~~P_+\psi^{(1)}=0\\
\label{zeromodes2}E=-Mc^2, \quad\to\quad  &\psi^{(1)}=0, ~~~~P_-\psi^{(2)}=0
\end{align}
\end{subequations}
The corresponding first order differential equations are readily solved explicitly.  Using Eq.~(\ref{scalar}) it can be shown that the acceptable (normalizable) solutions of Eq.~\eqref{zeromodes1} are given by
\begin{eqnarray}
\psi&=&Ne^{im\theta}\left(\ba{c}\varphi^{(1)}\\ 0 \ea\right),~~~~E=+Mc^2,\nonumber\\
\varphi^{(1)}&=&p^{m}\,(1-\b p^2)^{-\frac{1}{2\b\lam}}~~~~m\geq 0, ~~\lambda<0,
\end{eqnarray}
where $N$ is a normalization constant.

Similarly Eq.~\eqref{zeromodes2} is solved by:
\begin{eqnarray}
\psi&=&N\, e^{i(m+1) \vartheta}\,\left(\ba{c}0\\\varphi^{(2)}\ea\right),~~~~~E=-Mc^2,~~\phantom{xxxxx} \nonumber\\
\varphi^{(2)}&=&p^{-(m+1)}(1-\b p^2)^{\frac{1}{2\b\lam}}~~
%-\frac{1}{2\beta} < \lambda
~~m\leq -1,~~\lambda>0,
\end{eqnarray}
where $N$ is again a normalization constant. We then realize that when $\lambda<0$ there is a zero mode at $E=+Mc^2$ while for $\lambda >0$ the zero mode is in the negative branch of the spectrum. In both cases the zero mode is infinitely degenerate with respect to the angular quantum number $m$.

\section{Weak Magnetic Field (${\tilde{\omega}_c < \omega, \lam <0}$)}
\label{section:weak}
 Having discussed the zero modes explicitly in the previous section we now turn our attention to the general excited states, starting from the region of negative values of $\lambda$.\\
 
\noindent\underline{$(i)$ States of finite degeneracy.}\\
We start considering solutions $(a)$ in Table~\ref{combination1} and Table~\ref{combination2}. The  reduced eigenvalues $k_n^2$ (last column), for both components $\phi_{1,2}$ are obtained using the solutions of $A_1,B_1$, and $A_2,B_2$ (third  and fourth columns) in the eigenvalue relation in Eq.~(\ref{sol1}). In this case we have $A_1-B_1 = A_2-B_2 = -1/(2\beta\lambda)+m$ so that the eigenvalues of the two components match using for each one of them the same radial quantum number $n$. We must however take the intersection of the ranges of the angular quantum number $m$ from Table~\ref{combination1} and Table~\ref{combination2}. In this context we note that \[ \left\{\frac{1}{2\beta\lambda} < m \leq 0 \right\}
\cap \left\{\frac{1}{2\beta\lambda} < m \leq -1\right\}\]
\[\Longrightarrow \left\{\frac{1}{2\beta\lambda} < m \leq -1\right\} \]
Note also that this implies in particular  $ \frac{1}{2\beta \lambda} < -1$ which, since here $\lambda$ is negative, translates into a global condition $-\frac{1}{2\beta}<\lambda$ (i.e. valid for both components). All negative values $\lambda < -\frac{1}{2\beta}$ which would have been allowed for the solution of the upper component (Table~\ref{combination1}) must be excluded when matching with the solution of the lower component which requires $-\frac{1}{2\beta}<\lambda$.

 We then finally get:
\begin{equation}
\label{reduceddisc1}
\begin{array}{l}
k_{n,m}^2=-4 \left(-\frac{1}{2\beta\lambda}+m -n\right)^2~~~~, -\frac{1}{2\beta}<\lambda < 0\\  ~~~~~~~~n=1,\cdots < -\frac{1}{2\beta\lambda}+m, ~~~~~
\frac{1}{2\beta\lambda} < m \leq -1
\end{array}
\end{equation}
with the  eigenvalue $E_{n,m}$ of the Dirac problem  computed via Eqs.~(\ref{sol1},\ref{reducedeigenvalues}) and the corresponding upper and lower components of the spinor solution with the same radial quantum number $n$.\\ 

\noindent\underline{$(ii)$ States of infinite degeneracy.}\\
We now turn our focus on solutions $(b)$ of Table~\ref{combination1} and Table~\ref{combination2}.
The  reduced eigenvalues $k_n^2$ (last column), for both components $\phi_{1,2}$ are obtained again using the solutions of $A_1,B_1$, and $A_2,B_2$ (third  and fourth columns) in the eigenvalue relation in Eq.~\ref{sol1}. Now we find $A_1-B_1 = -\frac{1}{\beta\lambda}$ and $A_2-B_2 = -1/(\beta\lambda)-2$. We see then that the eigenvalues (last column) do not match if we use for both components the same radial quantum number $n$. This implies that in writing a spinor solution of the Dirac problem we must connect components with the radial quantum numbers shifted by one unit: e.g. $\varphi^{(1)}_{n}$ with $\varphi^{(2)}_{n-1}$. In addition we must take as acceptable only the intersection of the ranges of the angular quantum number $m$ from Table~\ref{combination1} and Table~\ref{combination2}. Thus: \[ \left\{ m \geq 0\right\}
\cap \left\{ m \geq -1 \right\}\Longrightarrow \left\{ m \geq 0 \right\}\] As in the case of the solutions $(a)$ while  
the condition in the third column of Table~\ref{combination1}  is always satisfied (being $\lambda <0$) the same condition on Table~\ref{combination2},  $ \frac{1}{2\beta \lambda} < -1$, implies $-\frac{1}{2\beta}<\lambda$  translates into a global condition  (i.e. valid for both components). Again, all negative values $\lambda < -\frac{1}{2\beta}$ which would have been allowed for the solution of the upper component (Table~\ref{combination1}) must be excluded when matching with the solution of the lower component which requires $-\frac{1}{2\beta}<\lambda$.

We then finally get:
\begin{equation}
\label{reduceddisc2}
\begin{array}{l}
k_{n,m}^2=-4 \left(-\frac{1}{2\beta\lambda}-n\right)^2~~~~, -\frac{1}{2\beta}<\lambda < 0\\  ~~~~~~~~n=1,\cdots < -\frac{1}{2\beta\lambda}+m, ~~~~~
 m \geq 0
\end{array}
\end{equation}
with the  eigenvalue $E_{n,m}$ of the Dirac problem  computed via Eqs.~(\ref{sol1},\ref{reducedeigenvalues}) and the corresponding upper and lower components of the spinor solution with the radial quantum number $n$ shifted by one unit, e.g. $\varphi^{(1)}_{n}$ with $\varphi^{(2)}_{n-1}$.

The full spinor solutions in the region $-\f{1}{2\b} <\lambda <0$, corresponding to cases $(a)$ (finite degeneracy, Eq.~\eqref{reduceddisc1}) and $(b)$ (infinite degeneracy, Eq.~\eqref{reduceddisc2}) illustrated in the above detailed discussion,  are explicitly written down in the following Eqs.~(\ref{negsamen},\ref{negdiffn}), in terms of unspecified  normalisation constants
$C_{-},C^{\prime}_{-}$.

\begin{widetext}
\beq
\label{negsamen}
\ba{l}
E_{n,m}=c\sqrt{M^2c^2+\f{1}{\b}-4\b\lam^2(-\f{1}{2\b\lam}+m-n)^2},~~~\psi_{n,m}=C_{-}\left(\ba{c} \displaystyle\f{\epsilon_{+}}{2m\lambda} e^{+im\vartheta}\varphi_{n,m}^{(1)}\vspace{0.2cm}\\ e^{+i(m+1)\vartheta}\varphi_{n,m}^{(2)}\ea\right),~~~n=1,....<-\f{1}{2\b\lam}+m\,, \\
\varphi_{n,m}^{(j)}= p^{B_j-\tfrac{1}{2}}(1-\b p^2)^{\tfrac{A_j-B_j}{2}}{}_2 F_1(-n,B_j-A_j+n;B_j+1/2;-\dfrac{\b p^2}{1-\b p^2}),~\f{1}{2\b\lam}< m\leq -1,~j=1,2\\
%y=2\, \text{sinh}^2(q)+1,~~~~A_1=-\f{1}{\b\lam}+\f{1}{2}+m,~~~~A_2=A_1-1,~~~~B_1=-m+\f{1}{2},~~~~B_2=B_1-1
\ea
\eeq
\beq
\label{negdiffn}
\ba{l}
E_{n,m}=c\sqrt{M^2c^2+\f{1}{\b}-4\b\lam^2(-\f{1}{2\b\lam}-n)^2},~~~~\psi_{n,m}=C_{-}^{\prime}\left(\ba{c}e^{+im\vartheta}\varphi_{n,m}^{(1)}\vspace{0.2cm}\\  e^{+i(m+1)\vartheta}\displaystyle\f{\epsilon_{-}}{2(m+1)\lam} \varphi_{n-1,m}^{(2)}\ea\right),~~~~n=1,....<-\f{1}{2\b\lam}\\
\varphi_{n,m}^{(j)}=p^{B_j-\tfrac{1}{2}}(1-\b p^2)^{\tfrac{A_j-B_j}{2}}{}_2 F_1(-n,B_j-A_j+n;B_j+1/2;-\dfrac{\b p^2}{1-\b p^2}),~~m\geq 0,~j=1,2
%\\ y=2\,\text{ sinh}^2(q)+1
%y=2\,\text{sinh}^2(q)+1,~~~~A_1=-\f{1}{\b\lam}+\f{1}{2}+m,~~~~A_2=A_1-1,~~~~B_1=m+\f{1}{2},~~~~B_2=B_1+1
\ea
\eeq

%\beq
%\label{negsamen}
%\ba{l}
%E_{n,m}=c\sqrt{M^2c^2+\f{1}{\b}-4\b\lam^2({\color{red}-}\f{1}{2\b\lam}+m-n)^2},~~~~\psi_{n,m}=\left(\ba{c}e^{+im\vartheta}\varphi_{n,m}^{(1)}\\ e^{+i(m+1)\vartheta}\varphi_{n,m}^{(2)}\ea\right),~~~~n=1,....<{\color{red}-}\f{1}{2\b\lam}+m\,, ~~~~ {\color{red}-\f{1}{2\b} <\lambda <0}\\
%\varphi_{n,m}^{(j)}= (\tanh(\f{\sinh^{-1}(\sqrt{\f{1+y}{2}}}{\sqrt{\b}}))^{-1/2} (y-1)^{B_j/2}(y+1)^{-A_j/2}~P_n^{(B_j-1/2,-A_j-1/2)}(y),~\f{1}{2\b\lam}< m\leq -1,~j=1,2\\
%%y=2\, \text{sinh}^2(q)+1,~~~~A_1=-\f{1}{\b\lam}+\f{1}{2}+m,~~~~A_2=A_1-1,~~~~B_1=-m+\f{1}{2},~~~~B_2=B_1-1
%\ea
%\eeq
%\beq
%\label{negdiffn}
%\ba{l}
%E_{n,m}=c\sqrt{M^2c^2+\f{1}{\b}-4\b\lam^2({\color{red}-}\f{1}{2\b\lam}-n)^2},~~~~\psi_{n,m}=\left(\ba{c}e^{+im\vartheta}\varphi_{n,m}^{(1)}\\ e^{+i(m+1)\vartheta}\varphi_{n-1,m}^{(2)}\ea\right),~~~~n=1,....<{\color{red}-}\f{1}{2\b\lam}\\
%\varphi_{n,m}^{(j)}=(\tanh(\f{\sinh^{-1}(\sqrt{\f{1+y}{2}}}{\sqrt{\b}}))^{-1/2}(y-1)^{B_j/2}(y+1)^{-A_j/2}~P_n^{(B_j-1/2,-A_j-1/2)}(y),~~m\geq 0,~j=1,2
%%\\ y=2\,\text{ sinh}^2(q)+1
%%y=2\,\text{sinh}^2(q)+1,~~~~A_1=-\f{1}{\b\lam}+\f{1}{2}+m,~~~~A_2=A_1-1,~~~~B_1=m+\f{1}{2},~~~~B_2=B_1+1
%\ea
%\eeq

\end{widetext}

\section{Strong Magnetic field (${\tilde{\omega}_c > \omega, \lam >0}$)}
\label{section:strong}
We now discuss the spectrum in the case when the magnetic field is such that the corresponding cyclotron frequency  $\tilde{\omega}_c $ is larger than the Dirac oscillator frequency $\omega$ which amounts to requiring $\lambda >0$. 

While the procedure is quite similar to the one exposed in the previous section the corresponding spectrum turns out to have important differences relative to the previous case. Such differences are responsible for the quantum phase transition(s). \\

%Since $\b>0,\lam>0$, the first of the above conditions always holds while the last one never satisfied irrespective of the magnitude or sign of $m$. However, the conditions $A>0, B>0$ require
%\beq
%m<\f{1}{\b\lam}-\f{1}{2},~~~~-\infty<m\leq -1
%\eeq
%Since the second of the above conditions ensures the first, 

\noindent\underline{$(i)$ States of infinite degeneracy.}\\
We now turn our focus on solutions $(c)$ of Table~\ref{combination1} and Table~\ref{combination2}.
The  reduced eigenvalues $k_n^2$ (last column), for both components $\phi_{1,2}$ are obtained again using the solutions of $A_1,B_1$, and $A_2,B_2$ (third  and fourth columns) in the eigenvalue relation in Eq.~\ref{sol1}. Now we have $A_1-B_1 = \frac{1}{\beta\lambda} -2$ and $A_2-B_2 = 1/(\beta\lambda)$. We see then that the eigenvalues (last column) do not match if we use for both components the same radial quantum number $n$. This implies that in writing a spinor solution of the Dirac problem we must connect components with the radial quantum numbers shifted by one unit: e.g. $\varphi^{(1)}_{n-1}$ with $\varphi^{(2)}_{n}$. Again we must take as acceptable only the intersection of the ranges of the angular quantum number $m$ from Table~\ref{combination1} and Table~\ref{combination2}. 
Thus: 
\[ \left\{ m \leq 0\right\}\,
\cap \, \left\{ m \leq -1 \right\}\,\Longrightarrow  \,\left\{ m \leq -1 \right\}\] As in the case of the solutions $(b)$ while  
the condition in the third column of Table~\ref{combination2}  is always satisfied (being $\lambda >0$) the same condition on Table~\ref{combination1},  $ \frac{1}{\beta \lambda} > 2$, implies $\lambda < \frac{1}{2\beta}$ and  translates into a global condition  (i.e. which must be valid for both components). Again, all positive values $\lambda > \frac{1}{2\beta}$ which would have been allowed for the solution of the lower component (Table~\ref{combination2}) must be excluded when matching with the solution of the upper component which requires $\lambda  < \frac{1}{2\beta}$.

We then finally get:
\beq\label{reduceddisc3}
\begin{array}{l}
k_{n,m}^2=-4(\f{1}{2\b\lam}-n)^2,\\n=1,....<\f{1}{2\b\lam},~~~~~~-\infty<m\leq -1\\
\end{array}
\eeq
with the  eigenvalue $E_{n,m}$ of the Dirac problem  computed via Eqs.~(\ref{sol1},\ref{reducedeigenvalues}) and the corresponding upper and lower components of the spinor solution with the radial quantum number $n$ shifted by one unit, e.g. $\varphi^{(1)}_{n-1}$ with $\varphi^{(2)}_{n}$.\\

%In this case the reduced eigenvalues can be found from (\ref{sol}) and are given by
%
%
%The second condition of Table~\ref{combination1} is satisfied subject to some constraints on $m$. Let us now assume that $\b\lam<0.5$ (which appears to be a reasonable assumption since $\b$ is a very small parameter). Then the second constraint is satisfied for all $m<0$. But the condition $B>0$ is satisfied if $m\geq -1$. So for the second case the energy eigenvalues and the corresponding admissible values of $m$ are given by
%\beq\label{e2}
%\begin{array}{l}
%%E_n=\pm c\sqrt{M^2c^2+\f{1}{\b}-4\b\lam^2(\f{1}{2\b\lam}-1-m-n)^2},\\~~~~n=0,1,....<\f{1}{2\b\lam}-1-m,~~~~~~-1\leq m<\f{1}{2\b\lam}-1\\ 
%k_{n,m}^2=-4(\f{1}{2\b\lam}-1-m-n)^2,\\~~~~n=0,1,....<\f{1}{2\b\lam}-1-m,~~~~~~-1\leq m<\f{1}{2\b\lam}-1\\ 
%\end{array}
%\end{equation}
%%Note that for $m=-1$, in (\ref{e1}) and (\ref{e2}) gives the same result. 

%It may be noted that the third condition is never satisfied for $m\leq 0$. But it is satisfied for very large positive values of $m$ :
%\beq
%\f{1}{2\b\lam}<m<\infty
%\eeq 
% However the conditions $A>0\Rightarrow m>\f{1}{\b\lam}+\f{1}{2}$ and $B>0\Ra -\infty<m\leq -1$ are clearly not compatible and therefore the third case of Table~\ref{combination1} is not admissible.
%xxxxxxxxxxxxxxxxxxxxxxxxxxxxxxxxxxxxxxxxxxxxxxxxxxxxxxxxxxxxxxxxxxxxxxxxxxxxxxxxxxxxxxxxxxxxxxxxxxxxxxxxxxxxxxxxxxxxxxxxxxxxxxx

\noindent\underline{$(ii)$ States of finite degeneracy.}\\
We are now left with the  solutions $(d)$ in Table~\ref{combination1} and Table~\ref{combination2}. The  reduced eigenvalues $k_n^2$ (last column), for both components $\phi_{1,2}$ are obtained using the solutions of $A_1,B_1$, and $A_2,B_2$ (third  and fourth columns) in the eigenvalue relation in Eq.~(\ref{sol1}). In this case we have $A_1-B_1 = A_2-B_2 = 1/(2\beta\lambda)-2-2m$ so that the eigenvalues of the two components match using for each one of them the same radial quantum number $n$. We must however take the intersection of the ranges of the angular quantum number $m$ from Table~\ref{combination1} and Table~\ref{combination2}. In this context we note that \[ \left\{0 \leq m < \frac{1}{2\beta\lambda} -1\,\right\}
\cap\, \left\{-1 \leq  m < \frac{1}{2\beta\lambda}-1\right\} \]\[\Longrightarrow \left\{0\leq  m< \frac{1}{2\beta\lambda}  -1 \right\}\]
Note also that this implies in particular  $ \frac{1}{2\beta \lambda} > 1$ which, since here $\lambda$ is positive, translates into a global condition $\lambda <\frac{1}{2\beta}$ (i.e. valid for both components). All positive values $\lambda > \frac{1}{2\beta}$ which would have been allowed for the solution of the lower component (Table~\ref{combination2}) must be excluded when matching with the solution of the upper component which requires $\lambda < \frac{1}{2\beta}$.

We then finally get:
\beq\label{reduceddisc4}
\begin{array}{l}
%E_n=\pm c\sqrt{M^2c^2+\f{1}{\b}-4\b\lam^2(\f{1}{2\b\lam}-1-m-n)^2},~~~~
%\\n=0,1,....<\f{1}{2\b\lam}-1-m,~~~~0\leq m<\f{1}{2\b\lam}-1\\
k_{n,m}^2=-4(\f{1}{2\b\lam}-1-m-n)^2,~~~~
\\n=1,....<\f{1}{2\b\lam}-1-m,~~~~0\leq m<\f{1}{2\b\lam}-1\\
\end{array}
\eeq
with the  eigenvalue $E_{n,m}$ of the Dirac problem  computed via Eqs.~(\ref{sol1},\ref{reducedeigenvalues}) and the corresponding upper and lower components of the spinor solution with the same radial quantum number $n$.

The full spinor solutions in the region $0 <\lambda <\f{1}{2\b}$ corresponding to cases $(c)$ (infinite degeneracy, Eq.~\eqref{reduceddisc3}) and $(d)$ (finite degeneracy, Eq.~\eqref{reduceddisc4}) illustrated in the above detailed discussion,  are explicitly written down in the following Eqs.~(\ref{possamen},\ref{posdiffn}) in terms of unspecified normalisation constants $C_{+},C^{\prime}_{+}$.
\begin{widetext}
\beq
\label{possamen}
\ba{l}
E_{n,m}=c\sqrt{M^2c^2+\f{1}{\b}-4\b\lam^2(\f{1}{2\b\lam}-n)^2},~~~~\psi_{n,m}=C_{+}\left(\ba{c}  e^{+im\vartheta} \displaystyle\f{\epsilon_{+}}{2m\lambda}\varphi_{n-1,m}^{(1)}\vspace{0.2cm}\\ e^{+i(m+1)\vartheta}\,\varphi_{n,m}^{(2)}\ea\right),~~~~n=1,....<\f{1}{2\b\lam}\\
\varphi_{n,m}^{(j)}=%e^{i(m-1+j)\vartheta} 
p^{B_j-\tfrac{1}{2}}(1-\b p^2)^{\tfrac{A_j-B_j}{2}}{}_2 F_1(-n,B_j-A_j+n;B_j+1/2;-\dfrac{\b p^2}{1-\b p^2}),~~-\infty<m\leq -1,~j=1,2
%\\y=2\,\text{sinh}^2(q)+1,~~~~A_1=\f{1}{\b\lam}-\f{3}{2}-m,~~~~A_2=A_1+1,~~~~B_1=-m+\f{1}{2},~~~~B_2=B_1-1
\ea
\eeq
\beq
\label{posdiffn}
\ba{l}
E_{n,m}=c\sqrt{M^2c^2+\f{1}{\b}-4\b\lam^2(\f{1}{2\b\lam}-1-m-n)^2},~~~~\psi_{n,m}=C_{+}^{\prime}\left(\ba{c}e^{+im\vartheta}\,\varphi_{n,m}^{(1)}\vspace{0.2cm}\\ \displaystyle\f{\epsilon_{-}}{2(m+1)\lam} e^{+i(m+1)\vartheta}\,\varphi_{n,m}^{(2)}\ea\right),~~~~n=1,....<\f{1}{2\b\lam}-1-m\\
\varphi_{n,m}^{(j)}=p^{B_j-\tfrac{1}{2}}(1-\b p^2)^{\tfrac{A_j-B_j}{2}}{}_2 F_1(-n,B_j-A_j+n;B_j+1/2;-\dfrac{\b p^2}{1-\b p^2}),~0\leq m<\f{1}{2\b\lam}-1,~j=1,2
%\\y=2\,\text{sinh}^2(q)+1,~~~~A_1=\f{1}{\b\lam}-\f{3}{2}-m,~~~~A_2=A_1+1,~~~~B_1=m+\f{1}{2},~~~~B_2=B_1+1
\ea
\eeq
%\beq
%\label{possamen}
%\ba{l}
%E_{n,m}=c\sqrt{M^2c^2+\f{1}{\b}-4\b\lam^2(\f{1}{2\b\lam}-n)^2},~~~~\psi_{n,m}=\left(\ba{c}e^{+im\vartheta}\,\varphi_{n-1,m}^{(1)}\\ e^{+i(m+1)\vartheta}\,\varphi_{n,m}^{(2)}\ea\right),~~~~n=1,....<\f{1}{2\b\lam}\\
%\varphi_{n,m}^{(j)}=%e^{i(m-1+j)\vartheta} 
%(\tanh(\f{\sinh^{-1}(\sqrt{\f{1+y}{2}}}{\sqrt{\b}}))^{-1/2}(y-1)^{B_j/2}(y+1)^{-A_j/2}~P_n^{(B_j-1/2,-A_j-1/2)}(y),~~-\infty<m\leq -1,~j=1,2
%%\\y=2\,\text{sinh}^2(q)+1,~~~~A_1=\f{1}{\b\lam}-\f{3}{2}-m,~~~~A_2=A_1+1,~~~~B_1=-m+\f{1}{2},~~~~B_2=B_1-1
%\ea
%\eeq
%\beq
%\label{posdiffn}
%\ba{l}
%E_{n,m}=c\sqrt{M^2c^2+\f{1}{\b}-4\b\lam^2(\f{1}{2\b\lam}-1-m-n)^2},~~~~\psi_{n,m}=\left(\ba{c}e^{+im\vartheta}\,\varphi_{n,m}^{(1)}\\ e^{+i(m+1)\vartheta}\,\varphi_{n,m}^{(2)}\ea\right),~~~~n=1,....<\f{1}{2\b\lam}-1-m\\
%\varphi_{n,m}^{(j)}=\left[\tanh\left(\f{\sinh^{-1}(\sqrt{\f{1+y}{2}}}{\sqrt{\b}}\right)\right]^{-\frac{1}{2}} (y-1)^{B_j/2}(y+1)^{-A_j/2}~P_n^{(B_j-1/2,-A_j-1/2)}(y),~0\leq m<\f{1}{2\b\lam}-1,~j=1,2
%%\\y=2\,\text{sinh}^2(q)+1,~~~~A_1=\f{1}{\b\lam}-\f{3}{2}-m,~~~~A_2=A_1+1,~~~~B_1=m+\f{1}{2},~~~~B_2=B_1+1
%\ea
%\eeq
\end{widetext}
%xxxxxxxxxxxxxxxxxxxxxxxxxxxxxxxxxxxxxxxxxxxxxxxxxxxxxxxxxxxxxxxxxxxxxxxxxxxxxxxxxxxxxxxxxxxxxxxxxxxxxxxxxxxxxxxxxxx
%%%%%%%%%%%%%%%%%%%%%%%%%%
%%%%V4
%\begin{table*}[t]
%\begin{ruledtabular} 
%\begin{tabular}{lccc}
%& $A$
%& $B$ 
%& $A>B$\\
%\hline
%$(a)$ 
%& $-\f{1}{\b\lam}-\f{1}{2}-m$
%& $-m-\f{1}{2}$ 
%& $ -\f{1}{\b\lam}>0$\\
%\hline
%$(b)$
%& $-\f{1}{\b\lam}-\f{1}{2}-m$
%& $m+\f{3}{2}$
%& $-\f{1}{\b\lam}-2-2m>0$\\
%\hline
%$(c)$
%& $ \f{1}{\b\lam}-\f{1}{2}+m$
%&$-m-\f{1}{2} $
%& $\f{1}{\b\lam}+2m>0$\\
%\hline
%$(d)$
%& $\f{1}{\b\lam}-\f{1}{2}+m  $
%& $m+\f{3}{2} $
%& $ \f{1}{\b\lam}-2>0$\\
%\end{tabular}
%\end{ruledtabular}
%\caption{{\color{red} here put some Table Caption ...slkfjalkfjlka}}
%\label{combination3}
%\end{table*}
%\begin{table*}[th]
%\begin{ruledtabular} \label{combination4}
%\begin{tabular}{lccc}
%& $A$
%& $B$ 
%& $A>B$\\
%\hline
%$(a)$ 
%& $-\f{1}{\b\lam}-\f{3}{2}-m$
%& $-m+\f{1}{2}$ 
%& $ -\f{1}{\b\lam}-2>0$\\
%\hline
%$(b)$
%& $-\f{1}{\b\lam}-\f{3}{2}-m $
%& $m+\f{1}{2}$
%& $-\f{1}{\b\lam}-2-2m>0$\\
%\hline
%$(c)$
%& $ \f{1}{\b\lam}+\f{1}{2}+m$
%&$-m+\f{1}{2} $
%& $\f{1}{\b\lam}+2m>0 $\\
%\hline
%$(d)$
%& $\f{1}{\b\lam}+\f{1}{2}+m  $
%& $ m+\f{1}{2} $
%& $ \f{1}{\b\lam}>0$\\
%\end{tabular}
%\end{ruledtabular}
%\caption{{\color{red} here put some Table Caption ...slkfjalkfjlka}}
%\label{combination4}
%\end{table*}

\begin{figure*}[ht!]
\includegraphics[width=15.75cm]{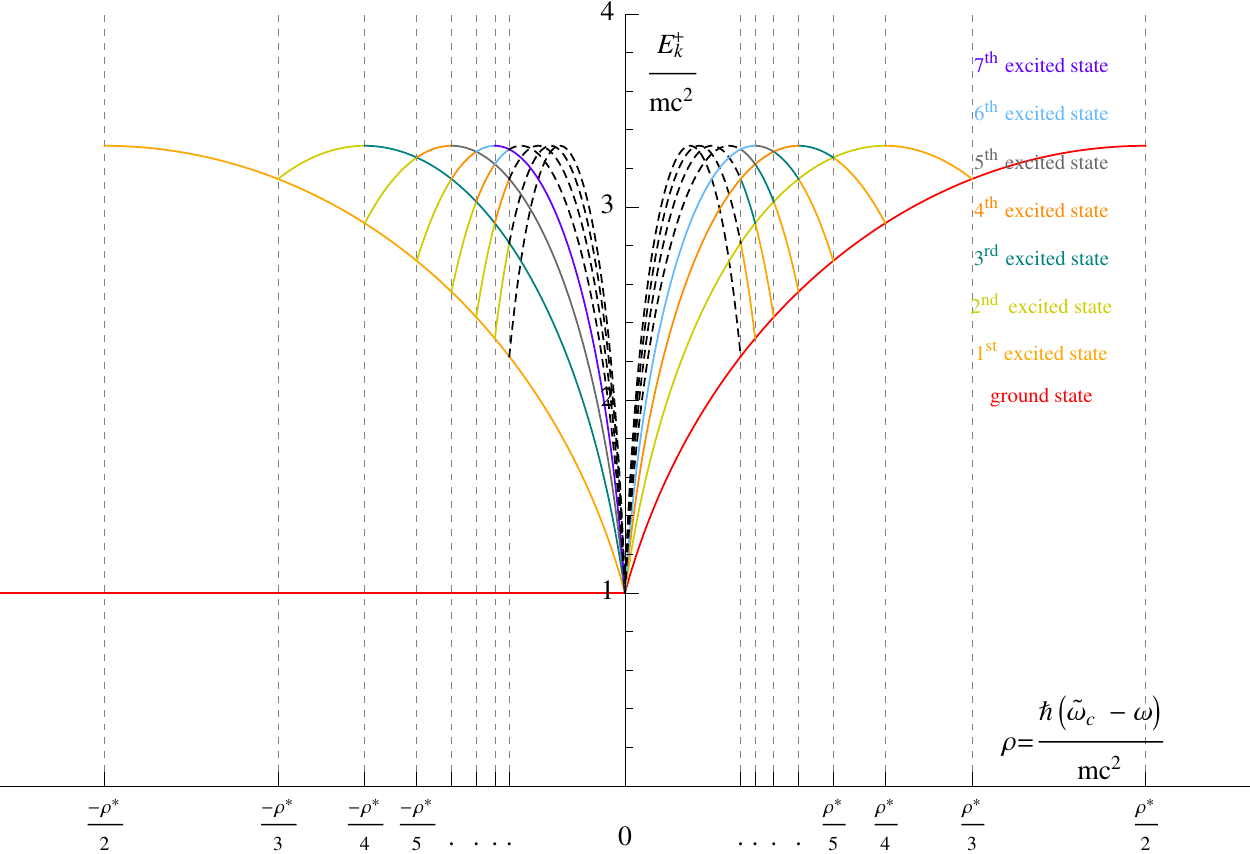}
\caption{\label{spectrum} (Color online) Positive branch of the discrete spectrum. The (positive) energy eigenvalues $E_{\cal N}$ from Eq.~\eqref{spectrumall}, in units of  $Mc^2$, are plotted as a function of the dimensionless variable $\rho$ which is directly related to the intensity of the external magnetic field $B$. Note that for a given value of the dimensionless variable $\rho$ we order the eigenvalues according to increasing values of the energy $E_{\cal N}$  resulting in the (excited) energy levels $E_k^+$ which are assigned a color code according to the legend.}
\end{figure*}
\begin{figure*}[ht!]
\includegraphics[width=15.75cm]{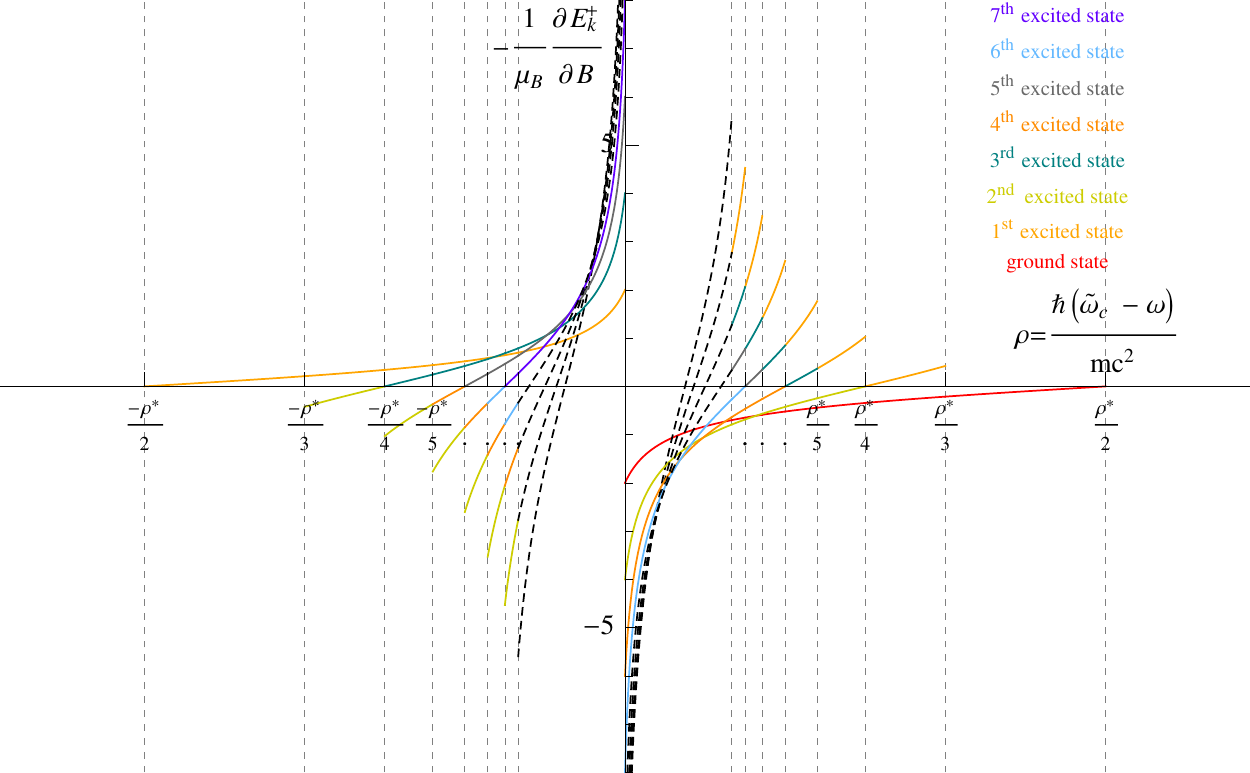}
\caption{\label{magnetisation}Magnetisation $M_{\cal N}=-\frac{\partial E_{\cal N}}{\partial B_0}$ of the system in units of the Bohr magneton. The color schemes is the same as in Fig.~\ref{spectrum} and follows the energy levels ordered by increasing numerical values of the energy eigenvalue. }
\end{figure*}
\section{Spectrum and Quantum Phase Transitions: Discussion}
\label{section:discussion}
It is well known that even in ordinary quantum mechanics ($\beta\to 0$ -- no maximum momentum --) when $\tilde{\omega}_c = \omega$ there is a quantum phase transition.
Here our main object is to study the impact of the Anti-Snyder model $\beta \neq 0$ with respect to the nature (and number) of such phase transition(s).

Let us consider for instance the spectrum $E_{n,m}$ as given in Eqs.~(\ref{negsamen},\ref{negdiffn}) and Eqs.~(\ref{possamen},\ref{posdiffn}). Upon introducing the dimensionless variable:
\begin{equation}
\rho=\frac{\hbar (\widetilde{\omega}_c-\omega)}{Mc^2}\,=\, \frac{\lambda}{M^2c^2},
\end{equation}
and the parameter:
\begin{equation}
\label{defrho}
\rho^* = \frac{1}{\beta M^2c^2}\,,
\end{equation}
the energy eigenvalues from Eqs.~(\ref{negsamen},\ref{negdiffn}) for $\lambda <0$, and in Eqs.~(\ref{possamen},\ref{posdiffn}) for $\lambda >0$  are easily obtained  as:
\begin{equation}
\label{spectrumall}
E_{\cal N} =\left\{\begin{array}{l}\pm Mc^2\, \sqrt{1-4\,{\cal N}\rho\left(1+\displaystyle\frac{\rho}{\rho*}\,{\cal N}\right)} \, , \quad \rho <0\\
\pm Mc^2\, \sqrt{1+4\,{\cal N}\rho\left(1-\displaystyle\frac{\rho}{\rho*}\,{\cal N}\right)}\, , \quad \rho >0\end{array} \right.
\end{equation}
where for $\rho <0$: ${\cal N} =n +|m|$ (negative $m$), Eq.~\eqref{negsamen}, and  ${\cal N} =n$, Eq.~\eqref{negdiffn}; and for $\rho >0$:
${\cal N} =n $, Eq.~\eqref{possamen}, and  ${\cal N} =n+m+1$ (positive $m$), Eq.~\eqref{posdiffn}.

%{\color{airforceblue} Note also that }

The energy eigenvalues $E_{\cal N}$ are plotted as function of the $\rho$ variable in Fig.~\ref{spectrum} where it is quite evident that the discrete spectrum is indeed bounded: for each value of $\rho$ there is a finite number of discrete energy levels and  there is a maximal energy corresponding to the maximum momentum of the Anti-Snyder model.

One first thing to notice about the spectrum is that it is always finite for any finite value of $\rho$ and it is always bounded from above $E\le E_{max}$ with $E_{max}$ easily computed from Eq.~\ref{spectrumall} as $E_{max}= Mc^2 \sqrt{1+\rho*}$ and straightforwardly related via Eq.~\ref{defrho} to the maximum momentum: $E_{max}=\sqrt{M^2c^4 + p_{max}^2c^2}$, as would be naturally expected for a relativistic particle with a maximum momentum.

For a given value of the dimensionless variable $\rho$ we order the eigenvalues according to increasing values of the energy $E_{\cal N}$  resulting in the (excited) energy levels $E_k^+$ which are assigned a color code according to the legend, see Fig.~\ref{spectrum}.
Since the functions $E_{\cal N}(\rho)$ from Eq.~\eqref{spectrumall}, are non monotonic the level suffix $k$ in general does not coincide with the suffix ${\cal N}$ which denotes instead a given function $E_{\cal N}(\rho)$. The energy levels are clearly separated in two distinct classes. One class of levels is built up (say for $\rho>0$) with the decreasing branches of the $E_{\cal N}(\rho)$ functions ($k=1,3,5,...$). This levels, which have the peculiar property that they disappear in the $\beta \to 0$ limit ($\rho^*\to \infty$), suffer a discontinuity at each value $\frac{\rho*}{2k+1+j}, \, j=1,2 \cdots$ which therefore denote a series of  quantum phase transitions accumulating towards the $\rho=0$ critical point. The class of levels (for $\rho>0$) built out of the increasing branches of the functions  $E_{\cal N}(\rho)$ (ground state, $k=2,4,6 \cdots)$ reduce in the  $\beta \to 0$ limit ($\rho^*\to \infty$) to the known standard result of ordinary quantum mechanics~\cite{Benitez:1990te} extending to $\rho \to \infty$.

Therefore we can give the following definition of the two classes of excited levels:
\begin{equation}
\label{classminus}
{\cal C}_{<} =\left\{\begin{array}{l}\, k=2,4,6, \cdots \, , \quad (\rho <0)\\
\,k=1,3,5, \cdots \, , \quad (\rho >0)\end{array} \right\}
\end{equation}
\begin{equation}
\label{clasplus}
{\cal C}_{>} =\left\{\begin{array}{l}\, k=1,3,5, \cdots \, , \quad (\rho <0)\\
\,k=2,4,6, \cdots \, , \quad (\rho >0)\end{array} \right\}
\end{equation}
We emphasise that in the vanishing $\beta$ limit  the class ${\cal C}_{>}$ reduces to the ordinary spectrum of the (2+1)-dimensional Dirac oscillator while the class ${\cal C}_{<}$ contains \emph{new levels} relative to the ordinary spectrum which disappear in the limit $\beta\to 0$ ($\rho^*\to \infty$) and thus:
\begin{equation}
\lim_{\beta\to 0}  {\cal C}_{<} = \{\emptyset\}\, .
\end{equation} 
Another interesting observable is the magnetisation of the system for a given energy level $M_{\cal N}=-\frac{\partial E_{\cal N}}{\partial B_0}$. The magnetisation $M_{\cal N}$ can be easily derived as:
\begin{equation}
M_{\cal N} = -\frac{\partial \rho}{\partial B_0} \, \frac{\partial E_{\cal N}}{\partial \rho}\, .
\end{equation}
From Eq.~\eqref{om} one has: $\frac{\partial \rho}{\partial B_0} = \frac{\hbar}{Mc^2} \left( \frac{\partial \tilde{\omega}_{c}}{\partial B_0}\right) =\frac{e\hbar}{2M^2c^3}$; and thus from Eq.~\eqref{spectrumall} the magnetisation  $M_{\cal N}$ can finally be cast as:
\begin{equation}
\label{magnetisationall}
M_{\cal N}\, =\,-\mu_B\,\left\{\begin{array}{l} \, \frac{\displaystyle -2{\cal N}-4\frac{\rho}{\rho^*}{\cal N}^2}{\displaystyle\sqrt{1-4\,{\cal N}\rho\left(1+\displaystyle\frac{\rho}{\rho*}\,{\cal N}\right)}} \, , \quad \rho <0\\
 \frac{\displaystyle2{\cal N}-4\frac{\rho}{\rho^*}{\cal N}^2}{\displaystyle\sqrt{1+4\,{\cal N}\rho\left(1-\displaystyle\frac{\rho}{\rho*}\,{\cal N}\right)}}\, , \quad \rho >0\end{array} \right.
\end{equation}
having introduced the Bohr magneton $\mu_B=\frac{e\hbar}{2Mc}$.

In Fig.~\ref{magnetisation} we plot the magnetisation $M_{\cal N}$ in units of the Bohr magneton as a function of $\rho$. 
It is evident that levels from the class ${\cal C}_{<}$ are characterised by a negative magnetisation when $\rho <0$ and by a positive magnetisation when $\rho >0$. Exactly the contrary applies to the class ${\cal C}_{>}$ so that in the limit $\beta \to 0$ one recovers the known result from ordinary quantum mechanics that the magnetisation is positive in the left chiral phase ($\rho <0$) while it is positive definite in the right chiral phase ($\rho >0$). The differences introduced by the new class of states in the Anti-Snyder model may have an impact in the termo-dynamical properties of such systems  and therefore may offer a way to access or at least constrain the maximum momentum parameter $\beta$.

Finally we briefly discuss  how the results presented in this work may have some relevance in the study of new materials like graphene~\cite{Geim:2007aa,Castro-Neto:2009aa,Semenoff:1984aa}, silicene~\cite{Lalmi:2010kx,Vogt:2012vn,Fleurence:2012ys,Lin:2012zr} and germanene~\cite{Davila:2014aa,Cahangirov:2009aa} in connection with possible experimental investigations of the AntiSnyder model.
Charge carriers in these materials are known to be described by an effective (2+1)-dimensional Dirac Equation.
In reference~\cite{Panella:2014hga} it was shown that both in graphene and silicene the quantum phase transitions induced by the non commutativity will affect differently the two inequivalent Dirac points $K$ and $K'$. Indeed the quantum phase transitions at $K$ and $K'$ arise for different critical  magnetic fields. A similar behaviour is worth investigating  here in the case of the AntiSnyder model, although one would have to cope with a multitude of critical fields as discussed above.  

  In view of the fact that in silicene and germanene  the charge carriers are massive the exact solutions derived in this work can be applied directly while in the case of graphene  the charge carriers are massless and  the same solutions  would need to be computed explicitly in the massless limit, $M\to0$.
   
\section{Conclusions}
\label{section:conclusion}
In this paper we have obtained exact solutions of the Dirac oscillator in the presence of a homogeneous magnetic field in the Anti-Snyder space. As mentioned before the same problem was examined in Snyder space (GUP with a minimal length) and it was shown that there were a number of quantum phase transitions~\cite{Menculini:2014isa} accumulating towards the $\rho =0$ critical point. Unlike the problem in ref.~\cite{Menculini:2014isa} in the present case the spectrum consists of a finite part as well as a continuum (scattering) part which will be discussed elsewhere.
Similarly to what has been found for the same system in the GUP model with a minimal length we find that the energy levels can be classified in two classes. The levels in one of these classes are such that they disappear in the vanishing $\beta$ limit (vanishing maximal momentum).  Each level in this class comes in alternatively in the spectrum ($k=2,4,6 \cdots$ when $\rho <0$ and $k=1,3,5 \cdots$ when $\rho >0$). They are all characterised by discontinuities  which are interpreted as quantum phase transitions. The levels in the other class instead when $\beta \to 0$ reduce to the known spectrum of the system in ordinary quantum mechanics, with a single critical point at $\rho=0$.

The magnetisation of the system is an observable that in principle has the potential to disentangle the states belonging to the class ${\cal C}_{<}$ from those of the class ${\cal C}_{>}$. Indeed we have found that the two classes are characterised by a magnetisation $M$ which is either positive definite or negative definite depending on $\rho$ being positive or negative. 

It is our belief that the results derived in this work  may prove useful for further studies of the Anti-Snyder model not only from the theoretical side but also from the experimental point of view given the recent progress in the physics of two dimensional systems involving materials like silicene, germanene and graphene whose charge carriers are effectively described by a (2+1)-dimensional Dirac equation.

\begin{acknowledgments}
This work is an outcome of the diploma thesis of M.~P. discussed in February 2015 at the University of Perugia, Department of Physics and Geology.

P.~R. acknowledges hospitality from the Physics Department of the University of Perugia and financial support from INFN - Istituto Nazionale di Fisica Nucleare - Sezione di Perugia.
\end{acknowledgments}

%\bibliography{AntiSnyder.bib}

%merlin.mbs apsrev4-1.bst 2010-07-25 4.21a (PWD, AO, DPC) hacked
%Control: key (0)
%Control: author (8) initials jnrlst
%Control: editor formatted (1) identically to author
%Control: production of article title (-1) disabled
%Control: page (0) single
%Control: year (1) truncated
%Control: production of eprint (0) enabled
%

\end{document}